\documentclass[letterpaper, 10 pt, conference]{IEEEtran}  

\IEEEoverridecommandlockouts                              

\usepackage[cmex10]{amsmath}
\usepackage{algorithmic}
\usepackage[font=footnotesize]{subfig}
\usepackage{graphicx}
\usepackage[xindy,acronym,nonumberlist]{glossaries}
\usepackage{hyperref}
\usepackage{textcomp}
\usepackage{arydshln}
\usepackage{tabularx}
\usepackage{booktabs}
\usepackage{nomencl} 
\usepackage{amsmath}
\usepackage{amssymb}
\usepackage[table]{xcolor}
\usepackage{soul,color}
\definecolor{grey}{rgb}{0.8,0.8,0.8}
\definecolor{lightgrey}{rgb}{0.9,0.9,0.9}
\usepackage{cite}

\usepackage{todonotes}

\hypersetup{colorlinks,urlcolor=blue, citecolor=blue, linkcolor=blue}

\newacronym{RFEC}{RFEC}{Remote Field Eddy Current}
\newacronym{RFT}{RFT}{Remote-Field Technologies}
\newacronym{MI}{MI}{Mutual Information}
\newacronym{PDF}{PDF}{Probability Density Function}
\newacronym{NDE}{NDE}{Non-Destructive Evaluation}
\newacronym{NDT}{NDT}{Non-Destructive Testing}
\newacronym{SVM}{SVM}{Support Vector Machine}
\newacronym{FEA}{FEA}{Finite Element Analysis}
\newacronym{ROR}{ROR}{Radius Outliers Removal}
\newacronym{SOR}{SOR}{Statistical Outliers Removal}
\newacronym{BandS}{B\&S}{Bell and Spigot}
\newacronym{SLAM}{SLAM}{Simultaneous Localisation and Mapping}
\newacronym{MFL}{MFL}{Magnetic Flux Leakage}
\newacronym{BEM}{BEM}{Eddy Current Broadband Electromagnetic}
\newacronym{SQUID}{SQUID}{Superconducting QUantum Interference Device}
\newacronym{LASSO}{LASSO}{Least Absolute Shrinkage and Selection Operation}
\newacronym{MSE}{MSE}{Mean Square Error}

\nomenclature{B}{magnetic field}
\nomenclature{$\omega$}{circular frequency}
\nomenclature{$\mu$}{magnetic permeability}
\nomenclature{$\sigma$}{electrical conductivity}
\nomenclature{$y$}{sensor measurement}
\nomenclature{$t$}{local pipe thickness}
\nomenclature{$w_i$}{model's parameter}
\nomenclature{$\boldsymbol{w}$}{set of all the model's parameter}
\nomenclature{$\boldsymbol{t}$}{thickness's matrix}

\makenomenclature

\begin{document}
\title{From the Skin-Depth Equation to the Inverse RFEC Sensor Model}

\author{Authors}
\author{Raphael~Falque, Teresa~Vidal-Calleja, Gamini Dissanayake, and Jaime~Valls~Miro\\
University of Technology Sydney, Australia \\ 
Emails: Raphael.H.Guenot-Falque@student.uts.edu.au, Teresa.VidalCalleja@uts.edu.au,\\
Gamini.Dissanayake@uts.edu.au, Jaime.VallsMiro@uts.edu.au}

\maketitle

\global\csname @topnum\endcsname 0
\global\csname @botnum\endcsname 0

\begin{abstract}
     In this paper, we tackle the direct and inverse problems for the Remote-Field Eddy-Current (RFEC) technology. The direct problem is the sensor model, where given the geometry the measurements are obtained. Conversely, the inverse problem is where the geometry needs to be estimated given the field measurements. These problems are particularly important in the field of Non-Destructive Testing (NDT) because they allow assessing the quality of the structure monitored. We solve the direct problem in a parametric fashion using Least Absolute Shrinkage and Selection Operation (LASSO). The proposed inverse model uses the parameters from the direct model to recover the thickness using least squares producing the optimal solution given the direct model. This study is restricted to the 2D axisymmetric scenario. Both, direct and inverse models, are validated using a Finite Element Analysis (FEA) environment with realistic pipe profiles.\\
\end{abstract}

\textbf{Keywords:} Remote Field Eddy Current (RFEC), direct problem, inverse problem, Non Destructive Evaluation (NDE)

\section{Introduction}
\label{sec:introduction}
\begin{figure}
        \centering
        \includegraphics[width=1\linewidth]{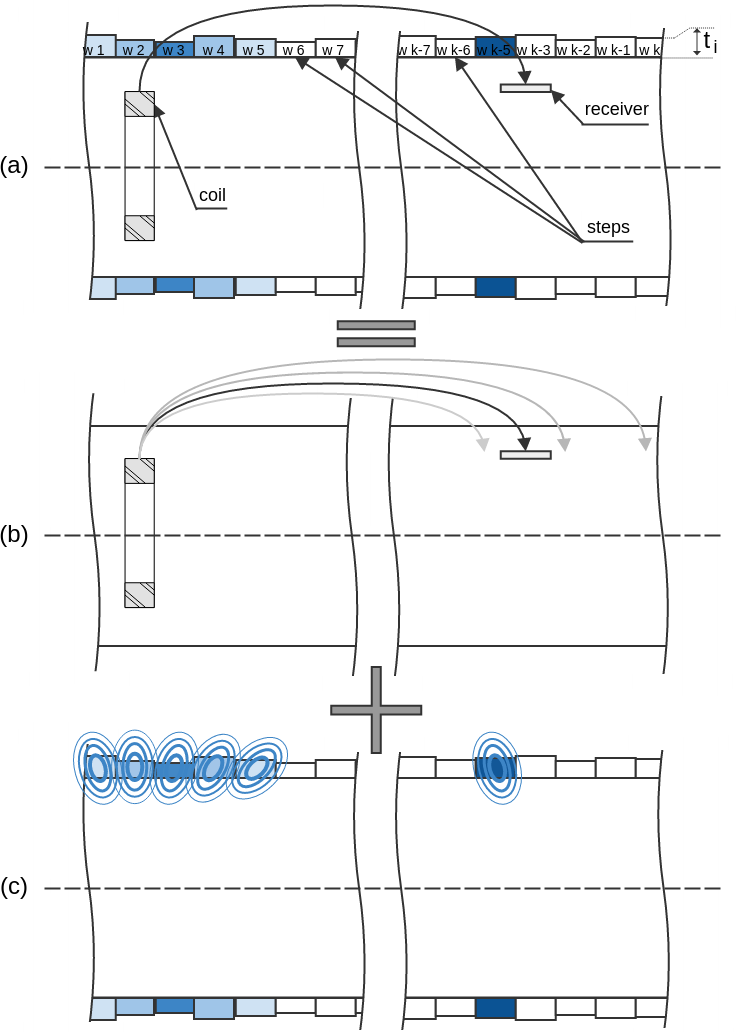}
        \caption{Representation of the RFEC phenomenon. From the global phenomenon (a), we propose a parametric direct model that consider independently the flow of the magnetic field in the air (b), and the local attenuation due to the magnetic field flowing through the pipe thickness (c).}
        \label{fig:pipe-geometry}
        \vspace{-0.1cm}
    \end{figure}

    The \gls{RFEC} technology allows in-line inspection of ferromagnetic pipelines. Tools based on this technology are usually composed of an exciter coil and one or several receivers. The exciter coil, driven by a low-frequency alternative current, generates an electromagnetic field that flows outside the pipe near the exciter coil and flows back inward the pipe at a remote area as shown in Fig.~\ref{fig:pipe-geometry}(a). The receivers are located in the remote part and record the magnetic field. As shown in the figure the magnetic field passes twice the pipe's wall; this phenomenon is commonly referred as the \textit{double through wall} in the literature~\cite{Atherton1995a}.
    
    When the magnetic field flows through a ferromagnetic medium (i.e. the pipe), the amplitude of the magnetic field is attenuated, and the phase is delayed. Due to the double through wall penetration, the magnetic field recorded by the receiver has been modified by different areas of the pipe: when it flows outward the pipe near the exciter coil and when it flows backwards the pipe in the remote area. Hence inferring the geometry of the pipe from the signal information is a challenging task since a single measurement is correlated with different areas of the geometry.

    Inferring the pipe's geometry from the tool signal corresponds to solving the inverse problem of the \gls{RFEC}. This problem has been studied in the literature for the 2D axisymmetrical case of a perfect pipe with a single crack. The problem is then formulated as recovering the shape (size and width) of the single defect~\cite{Davoust2006, Davoust2010, Tao2014}. These approaches solve the problem using data-driven techniques and bypass the problem of recovering the full pipe's geometry. These solutions fit the case of steel material, where most pipe bursts are due to cracks; in the case of cast-iron pipes, the material is more sensitive to corrosion. Hence the geometry of the pipe has a more organic shape rather than a single isolated crack. Therefore, for cast-iron pipes recovering the full pipe's geometry is critical.
    
    Some other approaches from the literature consist of modifying the tool design. The use of several receivers located at different axial locations from the exciter coil allows using redundancy of the information provided by passing through the same location to recover the full pipe's geometry~\cite{Cardelli1993,Skarlatos2008}. However, this approach leads to longer tools and require more electrical power to operate the multiple sensors and/or exciter coils. Due to the in-line nature of the \gls{RFEC} tools, the mobility and the battery consumption have to be optimised.
    
    Particularly for this work, to allow simple hardware design, we consider the case of an elementary \gls{RFEC} tool composed of a single exciter coil and a single receiver. The aim of this paper is to obtain an inverse sensor model of the \gls{RFEC} phenomenon which, given a set of continuous magnetic field measurements, allows to recover the full pipe's geometry for a 2D axisymmetric scenario.
    
    The remainder of the paper is organised as follows. In Sec.~\ref{sec:modelling}, we give conceptual ideas about the behaviour of the magnetic field. We then propose a direct model solved using \gls{LASSO}. From the direct model, we derive an inverse model formulated in a least-square form. The dataset generated with \gls{FEA} and the experimental results are given in Sec.~\ref{sec:results}. We finally discuss the performance and limitations of the proposed model in Sec.~\ref{sec:discussion}.
    
\section{Modelling of the RFEC phenomenon}
    \label{sec:modelling}
    
    The direct problem of the \gls{RFEC} phenomenon consists of mapping the pipe's geometry to the sensor measurement through a sensor model. Conversely, the inverse (or indirect) problem consists of finding the model that maps the sensor measurements into the pipe's geometry. The main goal is to solve the inverse problem, however, solving the direct problem provides qualitative and quantitative information on the form of the inverse model. Before to consider the direct and inverse problem, we discuss some high-level insight of the \gls{RFEC} technology. Particular attention is dedicated to understanding how the geometry near to the exciter coil impacts the sensor measurements. Qualitative descriptions of the overall \gls{RFEC} phenomenon have been broadly studied and in depth descriptions are available in the literature~\cite{Lord1988, Sun1996, Atherton1995a}.
    
    \subsection{Background information}
        As shown in~\cite{Atherton1995a}, it is possible to consider a defect in the pipe's geometry as an anomalous source model. The defect is then replaced by an independent source of magnetic field superposed to the pipe (see Fig. 12 in~\cite{Atherton1995a}). Knowing that the magnetic field gets attenuated while travelling through a ferromagnetic medium, the idea is to replace the lack of attenuation from the defect by a source of magnetic field superposed to a perfect pipe. Following the same idea, one could consider the pipe's thickness as an attenuation of the signal.
        
        Let us consider a pipe with an organic geometry (i.e. a corroded pipe), defined by a piecewise constant profile (as shown in Fig.~\ref{fig:pipe-geometry}). Each piece is then considered as a local source of attenuation. We then dissociate the global \gls{RFEC} phenomenon shown in Fig.~\ref{fig:pipe-geometry}(a) in two part: (i) the attenuation due to the magnetic field flowing in the air, and (ii) the local attenuation due to the magnetic field flowing through the pipe (the former one is shown in Fig.~\ref{fig:pipe-geometry}(b) and the latter in Fig.~\ref{fig:pipe-geometry}(c)).
        
        The global attenuation of the magnetic field propagating in the air (i), mostly due to field radiating from the coil, is a constant term for a given excitation and global geometry. The definition of this value is however complex since it involves many parameters (e.g. dimensions and excitation of the coil, diameter of the pipe, distance between the exciter and the receiver).
        
        The local interaction of the electromagnetic wave with the pipe (ii) can be described as a plane wave propagating through a homogeneous, isotropic, and conductive medium (i.e. the pipe). This phenomenon can be described by deriving the skin depth equation from the Maxwell equations and can be written as follow;
        \begin{equation}
            \text{B}(t) =\underbrace{\text{B}_0 e^{-\sqrt{\dfrac{\omega \mu \sigma}{2}}t}}_\text{amplitude}
            \underbrace{e^{-j\bigg(\sqrt{\dfrac{\omega \mu \sigma}{2}}t+\omega t\bigg)}}_\text{phase contribution},
            \label{eq:mf-propagation}
        \end{equation}
        with B the magnetic field, $\text{B}_0$ the initial value of the magnetic field, $\omega$ the frequency, $\mu$ the magnetic permeability of the medium, $\sigma$ the electrical conductivity, and $t$ the distance travelled by the wave. The amplitude and the phase-lag are usually the measurements recorded by the \gls{RFEC} tools since they have a log-linear or linear relationship with the thickness of the conductive medium:
        \begin{equation}
            \left\{
            \begin{array}{ll}
                \phi_{\text{local}} = \sqrt{\dfrac{\omega \mu \sigma}{2}}t\\
                ln(B)_{\text{local}} = ln(B_0) - \sqrt{\dfrac{\omega \mu \sigma}{2}}t
            \end{array}
            \right.
            \label{eq:skin-depth-equation}
        \end{equation}
        
        In this paper, we model the direct and inverse problem uniquely with the amplitude. However, a similar study could be done with the phase-lag.

    \subsection{Direct problem}
        We now consider the direct problem, which consists of finding a function $h$ such that $h:\boldsymbol{t}\longrightarrow y$, where $y$ is the sensor measurements and $\boldsymbol{t}$ a set of thickness values that describe the pipe's geometry around the \gls{RFEC} tool. 
        
        Let us first consider the case of a single measurement. Using the wave superposition principle, we can then add (i) and (ii) as follow;
        \begin{equation}
            y = y_0 - \sum_{i=1}^{k}w_i t_i + \epsilon,
            \label{eq:least-square-formulation}
        \end{equation}
        with $y_0$ the constant term described in (i), $t_i$ the $i^{th}$ thicknesses of the pipe piece (the pipe's geometry is approximated as a piecewise constant profile a shown in Fig.~\ref{fig:pipe-geometry}(a)), and $w_i$ the unknown parameters that embeds both $-\sqrt{\tfrac{\omega \mu \sigma}{2}}$ from (ii) and a location weight. Since this approach is an approximation of the actual phenomenon, we consider $\epsilon$ the noise contribution that contains both the actual sensor noise and an unmodelled non-linearity. Given enough independent measurements, the optimal values for the weights can be found using a least square formulation.
        
        Let us now consider a set of $m$ measurements where each measurement is associated to $k$ local average thicknesses that are regularly spaced over the length of the tool. This can be seen as moving the tool within the pipe simultaneously to gathering pipe thickness information in a sliding window. The sliding window approximates the geometry as a piecewise-constant profile as describe in Fig.~\ref{fig:pipe-geometry}(a). We then formulate Eq.~\eqref{eq:least-square-formulation} in a matrix form to combine the $m$ set of measurement and thickness values together,
        \begin{equation}
            \boldsymbol{y} = \boldsymbol{T} \boldsymbol{w} + \boldsymbol{\epsilon},
        \end{equation}
        
        The constant term $y_0$ from Eq.~\eqref{eq:least-square-formulation} is unknown (it depends on the excitation, the number of turn in the coil, the electromagnetic properties of the air and the distance between the exciter and the sensor). It is, however, possible to estimate $y_0$ from the measurements, therefore, we include it into the vector of the model parameters $\boldsymbol{w}$ which is defined as,
        \begin{equation}
            \boldsymbol{w} =     \begin{bmatrix}
               y_0\\
               w_1\\
               \vdots\\
               w_{k-1}\\
               w_k
                                    \end{bmatrix},
            \label{eq:parameters}
        \end{equation}
        $\boldsymbol{T}$ is the matrix that contains the local average thickness information;
        \begin{equation}
            \boldsymbol{T} = \begin{bmatrix}
                           1 & t_{11} & t_{12} & \dots  & t_{1k} \\
                           1 & t_{21} & t_{22} & \dots  & t_{2k} \\
                           \vdots & \vdots & \vdots & \ddots & \vdots \\
                           1 & t_{m1} & t_{m2} & \dots  & t_{mk}
                            \end{bmatrix},
        \end{equation}
        and $\boldsymbol{y}$ the vector with all the sensor measurements.
        \begin{equation}
            \boldsymbol{y} =    \begin{bmatrix}
                                        y_1\\
                                        y_2\\
                                        \vdots\\
                                        y_m
                                    \end{bmatrix}.
            \label{eq:y-definition}
        \end{equation}
        
        In order to select parameters $\boldsymbol{\hat{w}}$ that reflects the attenuation of the magnetic field through its path, there is a need for an optimisation method that sets the weights of the non-relevant thicknesses to zero. It can be obtained by learning the model parameters with \gls{LASSO}~\cite{Tibshirani1996}. Using this parameter selection also allows avoiding over-fitting the irrelevant parameter that would be performed by a closed form solution. More formally, \gls{LASSO} corresponds to the least square formulation with $L_1$-regularisation as
        \begin{equation}
            \text{min}\left[ \parallel \boldsymbol{T} \boldsymbol{\hat{w}} - \boldsymbol{y} \parallel^2 + \alpha \parallel \boldsymbol{\hat{w}} \parallel_1 \right],
            \label{eq:linear-least-square-direct-problem}
        \end{equation}
        with $\alpha$ the regularisation parameter, which is learned with an iterative process. Finally, the direct problem is solved by estimating $\hat{y}$ as
        \begin{equation}
            \boldsymbol{\hat{y}} = \underbrace{\boldsymbol{T} \boldsymbol{\hat{w}}}_{h(\boldsymbol{T})},
            \label{eq:y_hat-direct-problem}
        \end{equation}
        with the proposed model $h(\boldsymbol{T})$.
        
    \subsection{Inverse problem}
        After estimating the parameters of the direct model \eqref{eq:least-square-formulation}, we now consider the inverse problem. More formally, we want to find the inverse function $h^{-1}$ such that $h^{-1}:\boldsymbol{y}\longrightarrow \boldsymbol{t}$. Due to the double-through wall phenomenon, $h$ cannot be simply inverted as the geometry under the exciter coil and the receiver are convoluted in the measurements. Instead, having the direct problem expressed as a linear model, allows formulating the inverse problem in a closed form solution, which can be obtained with least squares. 
        
        We consider here solving the inverse problem for a long pipe section as one system (\emph{i.e.} recovering the thickness of the full pipe at the same time). To solve the optimisation problem through least squares, the degree of freedom (which is equal to the number of equations minus the number of parameters) of the system has to be positive or null. As a rule of thumb, to avoid over-fitting, the degree of freedom should be superior to ten. 
        
        Let us consider the inspection of a long pipe section using a \gls{RFEC} tool. During the inspection, a set of $m$ discrete measurements are collected at regular intervals along the pipe. We approximate the pipeline geometry as a piecewise-constant profile with $n$ steps of average thickness $t$. $n$ is chosen to be ten times smaller than $m$.
        
        We then re-write Eq.~\eqref{eq:least-square-formulation} so it can be formulated as a global optimisation problem, where all the sensor measurements $\boldsymbol{y}$ are related to all the piecewise thicknesses $t_i$ as 
        \begin{equation}
            \boldsymbol{y} = \boldsymbol{W} \boldsymbol{\hat{t}} + y_0,
        \end{equation}
        with $\boldsymbol{y}$ and $y_0$ defined in Eq.~\eqref{eq:least-square-formulation} and~\eqref{eq:y-definition}. $\boldsymbol{\hat{t}}$ is the set of the all the thickness estimates for each value of the piecewise-constant pipeline profile, defined as
        \begin{equation}
            \boldsymbol{\hat{t}} =    \begin{bmatrix}
                                        t_1\\
                                        t_2\\
                                        \vdots\\
                                        t_n
                                    \end{bmatrix}.
        \end{equation}
        
        $\boldsymbol{W}$ is an $m \times n$ matrix that contains the relationship between thickness values and sensor measurements and is defined by the parameters learned from the direct model. In practice, each line of $\boldsymbol{W}$ contains the weights $\boldsymbol{w}$ for the local thickness values and is set to $0$ for the others thickness values. Since there are multiple measurements between the $i^{th}$ and $(i+1)^{th}$ values, spatial weights $\overline{a_i}$ and $\overline{b_i}$ are used to define the influence of the piece proximity as:
        \begin{equation}
            \resizebox{0.87\hsize}{!}{%
            $
            \boldsymbol{W} = 
            \begin{bmatrix}
                \overline{a_1}w_1 & \overline{b_1}w_1 + \bar{a_1}w_2     & \dots        & \overline{b_1}w_k & 0 & 0 & \dots & 0\\
                
                \vdots     & \vdots                                     & \ddots     & \vdots & \vdots & \vdots & \ddots & \vdots \\
                
                \overline{a_j}w_1     & \overline{b_j}w_1 + \overline{a_j}w_2     & \dots     & \overline{b_j}w_k & 0 & 0 & \dots & 0\\
                
                0     & \overline{a_{j+1}}w_1     & \dots     & \overline{b_{j+1}}w_{k-1} + \overline{a_{j+1}}w_{k-1}     & \overline{b_{j+1}}w_k     & 0         & \dots & 0\\
                
                \vdots     & \vdots     & \ddots    & \vdots & \vdots & \vdots & \ddots & \vdots \\
                
                0     & 0     & \dots     & 0 & 0 & 0 & \dots & \overline{a_m}w_k\\
            \end{bmatrix},
            $}
        \end{equation}
        with $\overline{a_i}$ and $\overline{b_i}$ defined as follow,
        \begin{eqnarray}
            \bar{a_i} \triangleq \dfrac{b_i}{a_i+b_i} \\
            \bar{b_i} \triangleq \dfrac{a_i}{a_i+b_i}
        \end{eqnarray}
         where $a_i$ is the distance from the point measurement to the centre of the $i^{th}$ step, and $b_i$ the distance from the point measurement to the centre of the $(i+1)^{th}$ step. We then obtain the thickness estimates $\boldsymbol{t}$ by solving the linear least squares in closed form,
        
        \begin{equation}
            \boldsymbol{\hat{t}} = \underbrace{(\boldsymbol{W}^t\boldsymbol{W})^{-1} \boldsymbol{W}^t(\boldsymbol{y} - y_0)}_{h^{-1}(\boldsymbol{y})}.
            \label{eq:inverse_problem_least_square}
        \end{equation}

\section{Results}
    \label{sec:results}
    
    \gls{FEA} simulations with a 2D-axisymmetric geometry have been used to validate the proposed methods in a controlled environment. We look here at the performance of both the direct and inverse model applied to a long pipe section with a known geometry. Note that although the validation has been done for a 2D-axisymmetric scenario, the proposed models can be adapted for any \gls{RFEC} axisymmetric tool. 
    
    \subsection{FEA environment}
        \label{sec:FEA}
        
        \begin{figure}
            \centering
            \includegraphics[width=\linewidth]{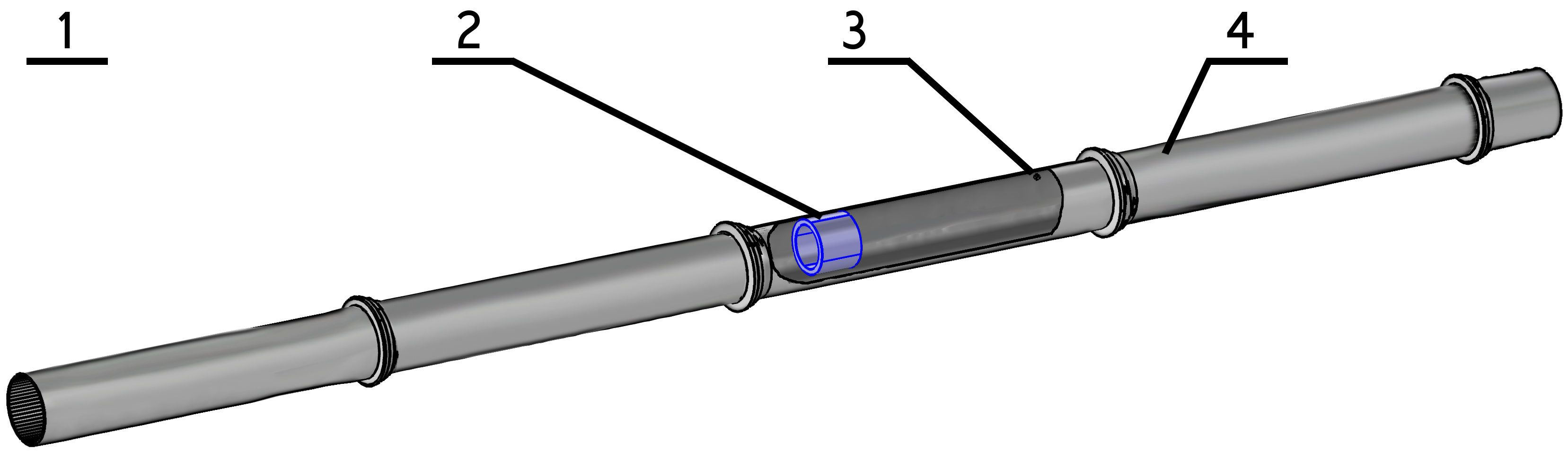}
            \caption{3D representation of the 2D axisymmetric simulation. (1) the air box, present all around the pipe, (2) the exciter coil as a rectangular cross multi-turn copper coil, (3) the receiver, simplified as a point measurement, and (4) the cast-iron pipe.}
            \label{fig:2D-axisymmetric}
        \end{figure}
        
        This section describes how the data for validation was obtained. In the context of our particular research project that motivates this paper, we used data from a cast-iron pipeline, which has been decommissioned and is currently dedicated for research purposes. This particular pipeline was laid more than hundred years ago, and some parts of the pipe are significantly corroded. Some pipes section have been exhumed, grid-blasted and analysed. The material properties of the cast-iron have been measured, and the corrosion's profile have been captured with a laser scanner, using the process described in~\cite{Skinner2014}. We generated a 60 m long 1-D profile based on the geometry of exhumed pipe segments. Once incorporated into a \gls{FEA} simulation environment, this realistic profile has provided sufficient data for validation.
        
        The \gls{FEA} used here is done using COMSOL Multiphysics in a 2D-axisymmetric scenario. The \gls{FEA} geometry is composed of four different components: (1) the air box defining the limits of the \gls{FEA} scenario, (2) the exciter coil, which is modelled as a rectangular cross multi-turn copper coil, (3) the receiver which is simplified to a point measurement (\emph{e.g.} it could simulate a hall effect sensor), and (4) the cast-iron pipe that has its geometry defined from pipe segments extracted from the decommissioned pipeline. A schematic of the global system is shown in Fig.~\ref{fig:2D-axisymmetric} with the thickness gains corresponding to \gls{BandS} joints that link pipe segments together.
        
        All medium are approximated as homogeneous and isotropic. The air and copper material properties are defined using built-in materials from the COMSOL library. To get a realistic 2D axisymmetric modelisation of the pipe, the pipe's magnetic properties are obtained by analysing a pipe sample with a \gls{SQUID}. We then have both the geometry and the material properties that come from a real pipeline. The material properties used in the model are displayed in Tab.~\ref{tab:materialMagneticProprieties}. The conductivity of the air is set to a non-zero value to avoid computational singularities. The stability of the simulation has been validated for different meshing sizes, air box sizes and other parameters.
        
        \begin{table}[b]
        \caption{Properties of each material}
        \centering
        \begin{tabular}{cllll}
            \hline
            material         & $\mu_r$     & $\epsilon_r$     & $\rho$ [S/m]    & l [m]        \\
            \hline
            Air                & 1            & 1                & 10            & 0.1        \\
            Cast-Iron        & 4.96        & 1                & 1.12e7        & 0.012        \\
            Copper (coil)    & 1            & 1                & 5.99e7        & 0.0115    \\
            \hline
        \end{tabular}
        \label{tab:materialMagneticProprieties}
        \end{table}
        
        The meshing size is defined according to the wavelength $\lambda$ of the magnetic field in each material (i.e. at least five times smaller than the wavelength), with $\lambda$ defined as:
        \begin{equation}
            \label{eq:lambda}
            \lambda = \dfrac{2\pi}{\sqrt{\dfrac{\omega.\mu.\sigma}{2}}}.
        \end{equation}        
    
        Using Eq.\eqref{eq:lambda} with the magnetic properties of each material we can define the minimum size for the meshing at each part of the scenario. The minimum size of each element in the meshing is given in Tab.~\ref{tab:materialMagneticProprieties}.

        The pipeline's inspection has been simulated using a parameter sweep for the position of the \gls{RFEC} tool within the pipe for the 60m length. The amplitude and the phase-lag of the electromagnetic phase have been recorded for each position of the parameter sweep.

    \subsection{Application of the direct model}
                    
        We now consider the direct problem applied to a dataset generated from the \gls{FEA} environment described previously. The aim here is to learn the parameters defined in Eq.~\eqref{eq:parameters}. As shown in Fig.~\ref{fig:2D-axisymmetric}, note that to make it more realistic the simulated thickness profile contains \gls{BandS} joints. The thickness of the \gls{BandS} joints are much larger than the other parts of the pipe, hence, due to the linear nature of proposed model, these data that relate to the~\gls{BandS} are expected to perform poorly. We solve the direct model for three datasets: (a) the first dataset include the complete set of data, (b) the data with a~\gls{BandS} joint located near the receiver have been removed in the second dataset, and (c) the data near both the exciter and receiver have been removed in the third dataset.
        
        \begin{figure*}
            \centering
            \subfloat[]{\label{fig:lls-dp-scatter-plots-full}\includegraphics[width=0.32\linewidth]{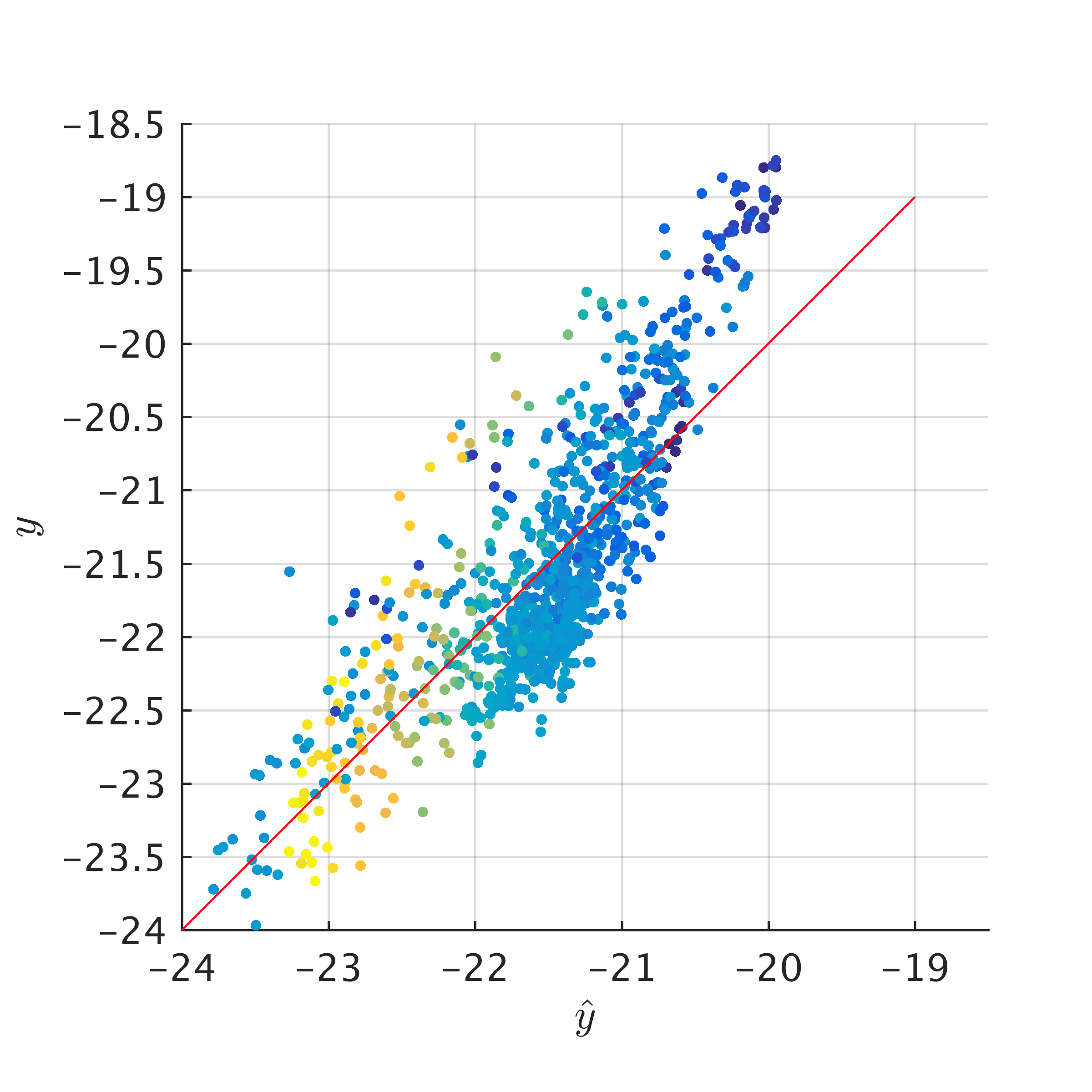}}\quad
            \subfloat[]{\label{fig:lls-dp-scatter-plots-partial}\includegraphics[width=0.32\linewidth]{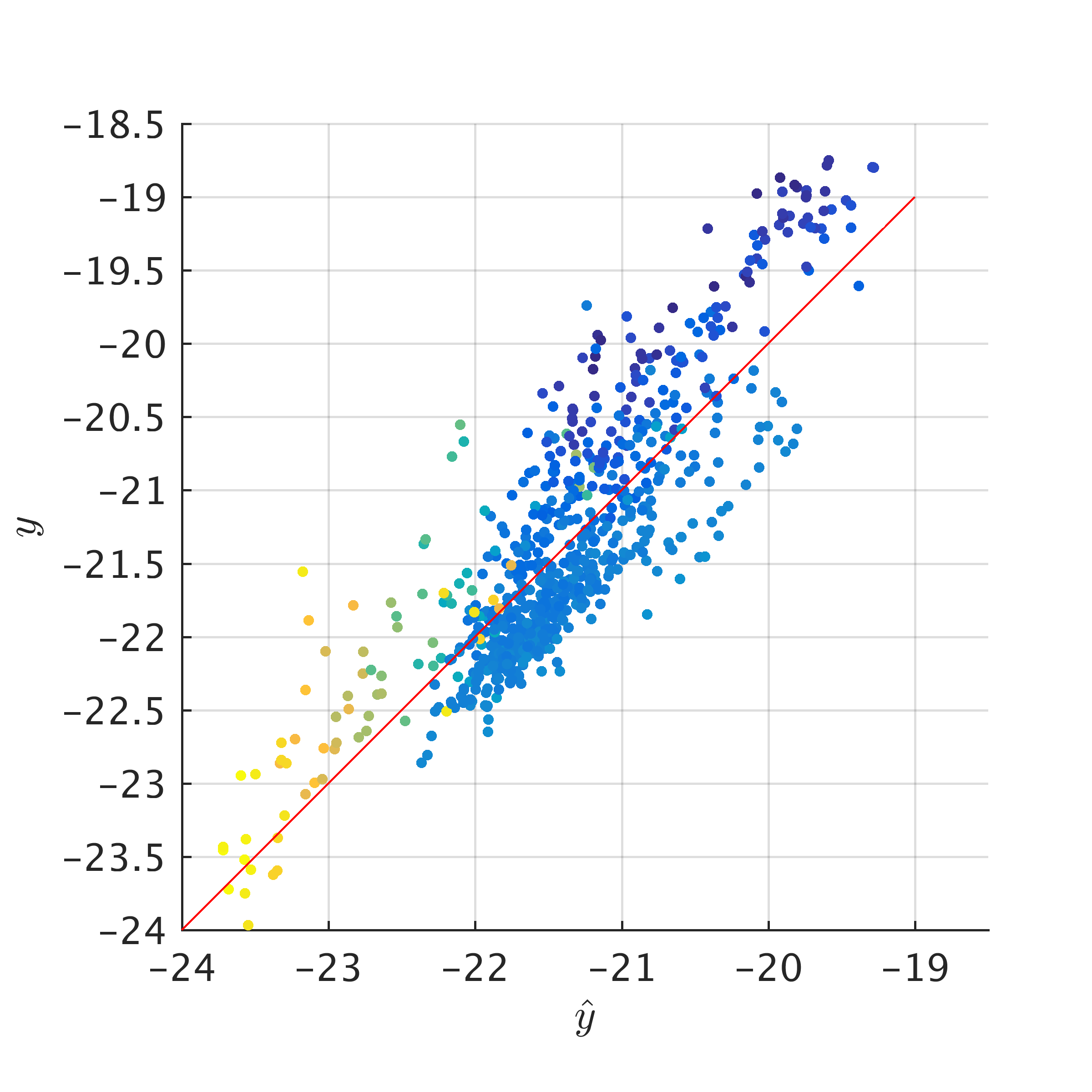}}\quad
            \subfloat[]{\label{fig:lls-dp-scatter-plots-minimal}\includegraphics[width=0.32\linewidth]{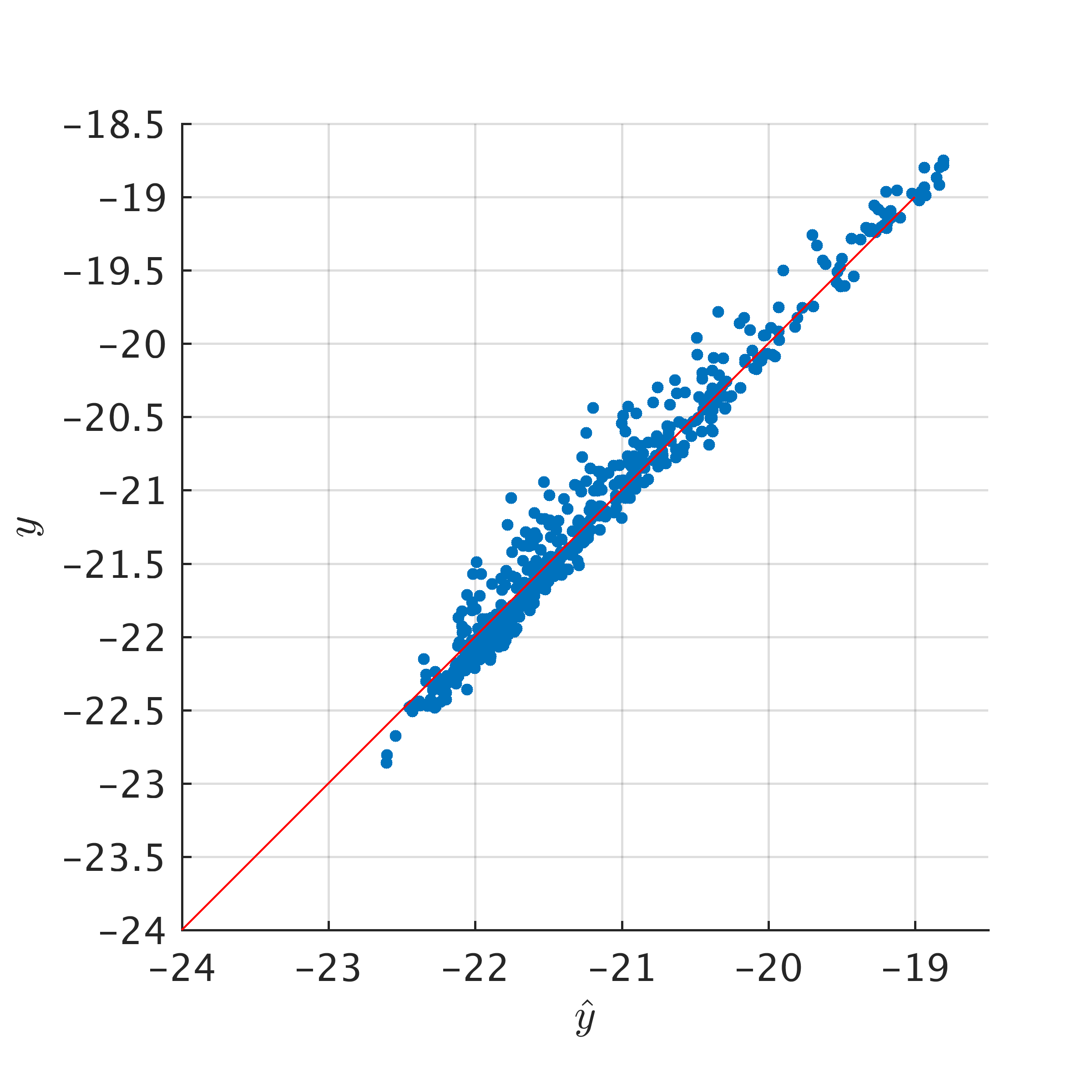}}
            
            \caption{Due to the presence of the \gls{BandS} joints inducing a sort of non-linearity in the data, the model described in Eq.~\eqref{eq:linear-least-square-direct-problem} is not longer valid. Therefore, we remove data where the \gls{BandS} joint has an impact on the exciter coil~\protect\subref{fig:lls-dp-scatter-plots-full}, and where it has an impact on the receivers~\protect\subref{fig:lls-dp-scatter-plots-partial}. The model learned from the filtered data is shown in~\protect\subref{fig:lls-dp-scatter-plots-minimal}.}
            \label{fig:lls-dp-scatter-plots}
        \end{figure*}

        We compared the estimated $\boldsymbol{\hat{y}}$ and the actual sensor measurement $\boldsymbol{y}$ in Fig.~\ref{fig:lls-dp-scatter-plots} with each sub-figure dedicated to each dataset. In Fig.~\ref{fig:lls-dp-scatter-plots}\protect\subref{fig:lls-dp-scatter-plots-full}, we set the colour information to reflect the impact of the~\gls{BandS} joints located near the receiver (the yellow points are more influenced by the \gls{BandS}). In Fig.~\ref{fig:lls-dp-scatter-plots}\protect\subref{fig:lls-dp-scatter-plots-full}, the blue points represent the estimation with the \gls{BandS} located on top of the receiver. The third dataset shown in Fig.~\ref{fig:lls-dp-scatter-plots}\protect\subref{fig:lls-dp-scatter-plots-minimal} shows a better regression.
        
        \begin{figure}[b!]
            \centering
            \includegraphics[width=\linewidth]{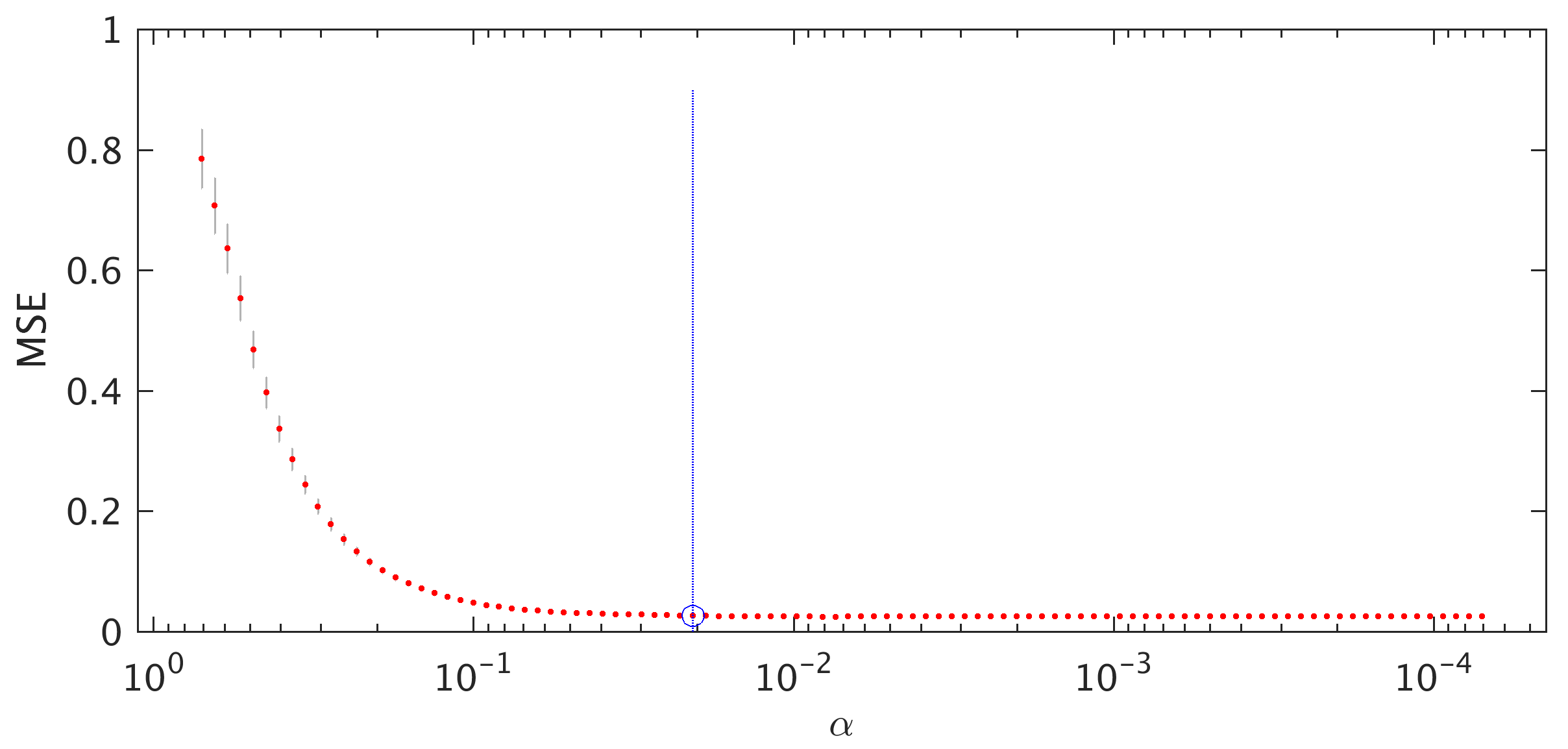}
            \caption{Evolution of the Mean Square Error (MSE) versus the value of the $\alpha$ parameter (using cross-validation). The $\alpha$ indicated in blue corresponds to the sparsest solution within one standard error of the MSE (it is the chosen one).}
            \label{fig:lasso}
        \end{figure}
        Since the simulations are done in a controlled environment, the locations of the \gls{BandS} joints are known, thus removing these particular data is a trivial task. In the case of an unknown environment, one could classify the construction features of the pipeline, which can be done using a \gls{SVM} classifier such as in~\cite{Vidal2014}. An alternative would consist of automating the data selection with methods such as Peirce's~\cite{Peirce1852} or Chauvenet's~\cite{Wiliam1863} criterion.
        
        The parameter $\alpha$ from Eq.~\eqref{eq:linear-least-square-direct-problem} is chosen using ten-fold cross-validation. The estimated parameters $\boldsymbol{w}$ and measurements of the goodness of fitting (Mean Square Error MSE and the coefficient of determination $R^2$) are available in Tab.~\ref{tab:LLS-weights}. As expected, the constant is a positive term, and the attenuation coefficients are negative terms. Moreover, we can see that the geometry near to the receiver and near to the exciter coil have a more important role which is reflected by higher weights.
        
        \begin{table*}
            \centering
            \caption{Output of the Least Absolute Shrinkage and Selection Operation. The localised increase of thickness (e.g. \gls{BandS}~joints) lead to the spread the weights. This is visible by comparing the lines of the table.}
            \begin{tabularx}{\textwidth}{ c | X | X c X X X | X X X X X X X X X | c c }
            
                & \textbf{cst} & \multicolumn{5}{c|}{\textbf{Exciter}} & \multicolumn{9}{c|}{\textbf{Receiver}} & \multicolumn{2}{c}{\textbf{Goodness of Fit}}\\
            
                \hline
                
                \textbf{coef} & $y_0$ & $w_1$ & $w_2$ & $w_3$ & $w_4$  & $w_5$  & $w_{19}$ & $w_{20}$  & $w_{21}$ & $w_{22}$  & $w_{23}$ & $w_{24}$  & $w_{25}$ & $w_{26}$  & $w_{27}$& MSE & $R^2$\\
            
                \hline
            
                \textbf{dataset (a)} & \cellcolor{grey}\tiny-17.6 & \cellcolor{grey}\tiny-26.0 & \cellcolor{grey}\tiny-20.9 & \cellcolor{grey}\tiny-5.7 & \cellcolor{grey}\tiny-5.7 & \cellcolor{lightgrey}\tiny-3.2 & \cellcolor{lightgrey}\tiny-0.2 & \cellcolor{lightgrey}\tiny-2.0 & \cellcolor{lightgrey}\tiny-2.7 & \cellcolor{grey}\tiny-7.8 & \cellcolor{grey}\tiny-10.3 & \cellcolor{grey}\tiny-19.6 & \cellcolor{grey}\tiny-19.1 & \cellcolor{grey}\tiny-8.6 & \cellcolor{grey}\tiny-9.1 & 0.296 & 0.5754\\
            
                \textbf{dataset (b)} &\cellcolor{grey}\tiny -16.7 &\cellcolor{grey}\tiny  -22.4 &\cellcolor{grey}\tiny  -19.9 &\cellcolor{lightgrey}\tiny   -3.3 &\cellcolor{lightgrey}\tiny   -5.5 &\cellcolor{lightgrey}\tiny   -1.9 &\cellcolor{lightgrey}\tiny   -1.1 &\cellcolor{lightgrey}\tiny   -1.6 &\tiny      0 &\cellcolor{lightgrey}\tiny   -6.2 &\cellcolor{grey}\tiny  -41.5 &\cellcolor{grey}\tiny  -24.5 &\cellcolor{grey}\tiny  -63.1 &\cellcolor{lightgrey}\tiny   -2.5 &\tiny      0 & 0.174 & 0.6391\\
            
                \textbf{dataset (c)} &\cellcolor{grey}\tiny   -16.0 &\cellcolor{grey}\tiny  -48.3 &\cellcolor{grey}\tiny -103.0 &\tiny      0 &\tiny      0 &\tiny      0 &\tiny      0 &\tiny      0 &\tiny      0 &\cellcolor{lightgrey}\tiny   -2.9 &\cellcolor{lightgrey}\tiny  -11.2 &\cellcolor{grey}\tiny  -46.3 &\cellcolor{grey}\tiny  -36.9 &\cellcolor{grey}\tiny  -29.0 &\cellcolor{lightgrey}\tiny   -3.9 & 0.016 & 0.8779\\
            \end{tabularx}
            \label{tab:LLS-weights}
            \vspace{-0.2cm}
        \end{table*}
        
    \subsection{Application of the inverse problem}
        After solving the direct problem, all the parameters required for the inverse problem are known. We consider recovering the 60 metres of pipe thickness as a global problem and the full geometry is recovered from the set of all the measurement using the formulation established in Eq.~\eqref{eq:inverse_problem_least_square}.
        
        The inverse problem relies on the parameters learnt for the direct problem. In the case where all the parameters from the~\gls{RFEC} tool and the magnetic properties of the pipe specimen are known, it is possible to obtain direct and inverse models through \gls{FEA} simulation. Otherwise, multiple thickness measurements have to be collected from the studied pipe. These thickness measurements at specific locations are needed to learn the parameters. In practice, collecting such measurements is a feasible task considering the few parameters that are present in the proposed model.
        
\begin{figure}[b]
            \centering
            \includegraphics[width=\linewidth]{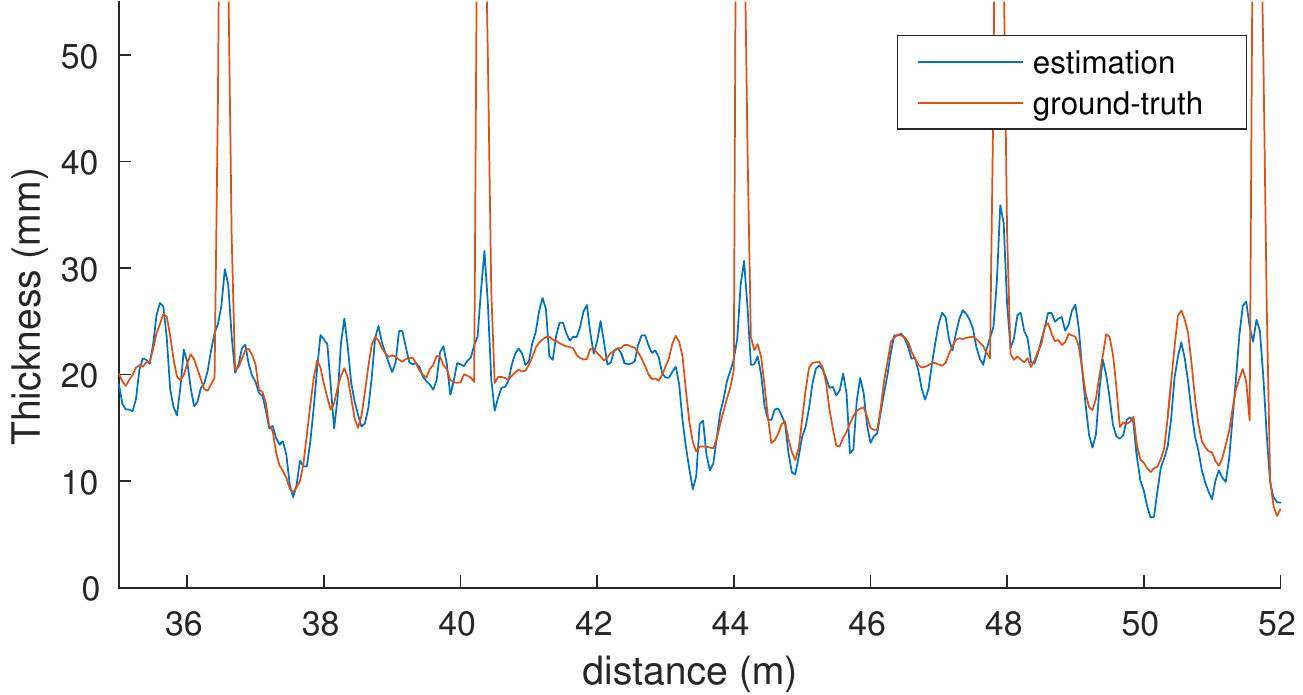}
            \caption{Thickness estimated in closed-form. The estimation is shown in blue, and the ground-truth  in orange.}
            \label{fig:indirect-problem-results}
        \end{figure}

        A zoom-in of the pipe profile reconstructed is shown in Fig.~\ref{fig:indirect-problem-results}. The estimation is shown in blue, and the ground truth is shown in orange. The spikes in the ground-truth correspond to the~\gls{BandS} joints. As predicted, the proposed inverse model cannot recover these thicknesses due to the non-linear behaviour of the magnetic field in these regions.
        
        The estimation error for the thickness of the 60 metre-long pipe is of 0.1mm for the MSE and 10.3mm for RMSE (with the average thickness of the pipe being around 30mm). If we remove the areas with the \gls{BandS} joints, the RMSE falls to 2.6mm.

\section{Discussion}
    \label{sec:discussion}
    In this paper, we tackle the direct and inverse problems for a 2D-axisymmetric RFEC tool composed of a single exciter coil and a single receiver. We have shown, using~\gls{FEA}, that both direct and inverse model are accurate for recovering pipe sections with organic geometry (which is often the case for corroded cast-iron pipes). The FEA model used to generate the dataset is based on a realistic geometry and material properties obtained from old cast-iron pipes.
    
    The proposed direct model is solved using LASSO. The $L_1$-regularisation allows selecting automatically the important thickness areas for the model while reducing the number of parameters. This result into a simplistic model, with most important thicknesses located next to the exciter coil and the receiver. The inverse problem relies on the parameters from the direct problem and is solved using least squares. For training the proposed inverse model, thickness measurements have to be collected from the pipe. In practice, collecting such measurements is a feasible task considering the few parameters of the proposed model.
    
    The main limitation of the proposed method lies in the form of the proposed model. The linear model allows solving the inverse problem in a closed-form. The model gives accurate results apart for the \gls{BandS} joints. For these extremely thick thicknesses, the magnetic field would flow through the path of least resistance which cannot be captured by a linear model. Furthermore, to the outstanding thicknesses, the magnetic properties are considered constant for the full pipeline. In practice, pipes can have a variation of magnetic properties; this case is not studied here.
    
    In future work, we are planning to apply this method for a tool with a sensor array (3D case). It can be shown that the attenuation from the exciter behaves as a circumferential offset~\cite{Falque2014}. Therefore, it is possible to deconvolute the signal in a similar fashion.
    
\section*{Acknowledgment}
    This publication is an outcome from the Critical Pipes Project funded by Sydney Water Corporation, Water Research Foundation of the USA, Melbourne Water, Water Corporation (WA), UK Water Industry Research Ltd,  South Australia Water Corporation, South East Water, Hunter Water Corporation, City West Water, Monash University, University of Technology Sydney and University of Newcastle. The research partners are Monash University (lead), University of Technology Sydney and University of Newcastle.

\bibliography{From_the_Skin-Depth_Equation_to_the_Inverse_RFEC_Sensor_Model}
\bibliographystyle{IEEEtranNoURL}
\end{document}